\documentclass{myPoS}

\usepackage{lineno}

\title{Precision Measurements in the Higgs Sector at ATLAS and CMS}

\ShortTitle{Precision Measurements in the Higgs Sector~~}

\author{\speaker{Andr\'e Sopczak on behalf of the ATLAS and CMS Collaborations}\\
        IEAP CTU in Prague
\\
        E-mail: \email{andre.sopczak@cern.ch}}

\abstract{
A concise review of precision measurements in the Higgs sector of the Standard Model (SM) 
of particle physics is given using ATLAS and CMS data.
The results are based on LHC Run-2 data, taken between 2015 and 2018.
Impressive progress has been made since the discovery of the Higgs boson in 2012 
for measuring all major production and decay modes. 
Good agreement with the SM predictions was observed in all measurements.
}

\FullConference{International Conference on Precision Physics and Fundamental Physical Constants - FFK2019\\
		9-14 June, 2019\\
		Tihany, Hungary}

\begin{document}

\section{Introduction}
The Higgs boson of the SM is produced at the Large Hadron Collider (LHC) at CERN in different production modes. They are given below together with the production cross-sections~\cite{deFlorian:2227475} and the numbers of expected Higgs boson events for 140\,fb$^{-1}$ corresponding to the complete Run-2 data set:
\begin{itemize}
\item gluon-gluon fusion (ggF), 48.58\,pb, 6.80M events,
\item vector boson fusion (VBF), 3.783\,pb, 530k events,
\item vector boson associated production (WH+ZH), $(1.415+0.8915)$\,pb, 323k events, and
\item top-top Higgs (ttH) production, 0.507\,pb, 71k events.
\end{itemize}

The decay branching ratios of the Higgs boson in the SM are given in Table~\ref{tab:br}.

\begin{table}[htbp]
\caption{SM Higgs boson decay branching ratios.
\vspace*{-3mm}
}
\label{tab:precision}
\begin{center}
\renewcommand{\arraystretch}{1.0}
             \begin{tabular}{c|cccccc}
                        \hline\hline
$H\rightarrow$&$bb$&$WW^*$&$\tau\tau$&$ZZ^*$&$\gamma\gamma$&$\mu\mu$ \\ \hline
B    (\%)             &58    & 21          & 6.3            & 2.6      &     0.23                   &  0.022\\ 
 \hline\hline
\end{tabular}
\end{center}
\label{tab:br}
\end{table}

ATLAS~\cite{Aad:1129811} and CMS~\cite{Chatrchyan:1129810} measured the Higgs boson properties as 
outlined in the following sections.

\section{Higgs boson mass and width}
The Higgs boson mass has been determined by 
ATLAS $m_H = 124.97 \pm 0.24$\,GeV~\cite{Aaboud:2621113} and 
CMS $m_H = 125.35 \pm 0.15$\,GeV~\cite{CMS-PAS-HIG-19-004}.
Indirect limits on the Higgs boson width are set at 95\% CL by 
ATLAS $\Gamma_H<14.4$\,MeV~\cite{Aaboud:2633244} and
CMS $\Gamma_H<9.16$\,MeV~\cite{Sirunyan:2652808}.

\section{Higgs boson couplings to bosons}

\subsection{$H\rightarrow \gamma\gamma$}
ATLAS and CMS measured the coupling $H\rightarrow \gamma\gamma$. 
ATLAS obtains for 79.8 fb$^{-1}$ data, ggF, VBF, VH and ttH production a $\sigma \times B$ value 
consistent with the SM expectation~\cite{ATLAS-CONF-2018-018}.
Using 36.9\,fb$^{-1}$ data, CMS measured the total Higgs boson production
cross-section
$\sigma_{\rm tot} = 61.1 \pm 6.0~({\rm stat}) \pm 3.7~({\rm syst})$\,pb
based on a combination of the
$H\rightarrow \gamma\gamma$ and 
$H\rightarrow ZZ$ channels,
which is consistent with the SM value 
$\sigma_{\rm tot}^{\rm SM} = 55.6 \pm 2.5$\,pb~\cite{Sirunyan:2651932}.

\subsection{$H\rightarrow ZZ$}
A clear signal over background is observed for $H\rightarrow ZZ \rightarrow 4l$ for 79.8\,fb$^{-1}$ ATLAS data~\cite{ATLAS-CONF-2018-018} 
and 137.1\,fb$^{-1}$ CMS data~\cite{CMS-PAS-HIG-19-001}.
For the ggF production, the $\sigma \times B$ measurement has an uncertainty of about 15\%.
The measurement is consistent with the SM expectation.

\subsection{$H\rightarrow WW$}
\vspace*{-1.5mm}
In the ATLAS $H\rightarrow WW$ analyses, the following results are obtained for ggF production $\sigma \times B = 11.4^{+1.2}_{-1.1}\,({\rm stat})\,_{-1.1}^{+1.2}\,({\rm theo,syst})\,_{-1.3}^{+1.4}\,({\rm exp,syst})$\,pb, which is
consistent with the SM expectation $10.4 \pm 0.6$\,pb. 
For VBF  production $\sigma \times B =0.5_{-0.22}^{+0.24}\,({\rm stat}) \pm 0.10\,({\rm theo,syst})\,_{-0.13}^{+0.12}\,({\rm exp,syst})$\,pb, which is also consistent with the SM expectation of $0.81\pm 0.02$\,pb~\cite{Aaboud:2636382}.

\vspace*{-3mm}
\section{Higgs boson couplings to fermions}
\vspace*{-1mm}
\subsection{$H\rightarrow \tau\tau$}
\vspace*{-1.5mm}
In combined ATLAS and CMS data (LHC Run-1), the observation (expectation)\footnote{The expected significance is given in parenthesis throughout the article.} was 5.5 (5.0) 
st.\,dev.\,\cite{Aad:2158863}.\,ATLAS Run-1 and 36 fb$^{-1}$\,Run-2 data led to 6.4 (5.4) 
st.\,dev.\,\cite{Aaboud:2648520}.
In\,ggF\,production, the measurement is $3.1 \pm 0.1\,({\rm stat})\,_{-0.13}^{+0.16}\,({\rm syst})$\,pb, compared to the SM expectation of $3.05 \pm 0.13$\,pb.
For VBF production, the measurement is $0.28 \pm 0.09\,({\rm stat})\,_{-0.09}^{+0.11}\,({\rm syst})$\,pb, compared to the SM value $0.237 \pm 0.006$\,pb. 
Similar results are obtained by CMS for 35.9\,fb$^{-1}$ data~\cite{Sirunyan:2276465}, 
leading to 4.9 (4.7) st. dev. and 5.9 (5.9) st. dev. when combined with Run-1 data.

Stage-1 simplified template cross-sections (STXS)~\cite{CMS-PAS-HIG-18-032} are based on about 77.4\,fb$^{-1}$  CMS data, taken 2015-2017: 
$\sigma(pp\rightarrow H) \times B(H\rightarrow \tau\tau) = 2.56 \pm 0.48\,({\rm stat}) \pm 0.34\,({\rm syst})$\,pb,\\
$\sigma(gg\rightarrow H, bbH)\times B(H\rightarrow \tau\tau)=1.11\pm0.81\,({\rm stat})\pm0.78\,({\rm syst})$\,pb, and\\ 
$\sigma({\rm VBF})\times B(H\rightarrow\tau\tau) = 0.34 \pm 0.08\,({\rm stat}) \pm 0.09\,({\rm syst})$\,pb.

The CP invariance of the Higgs boson coupling to vector bosons has been tested in the VBF
$H\rightarrow \tau\tau$ process in 36.1\,fb$^{-1}$ ATLAS data.
No evidence of CP violation is observed,
consistent with the SM expectation~\cite{ATLAS-CONF-2019-050}.
 
\vspace*{-1mm}
\subsection{$H\rightarrow bb$}
\vspace*{-1.5mm}
This is a difficult channel due to large backgrounds, despite the large branching ratio (58\%).
The most sensitive production mode is VH, and
combined ATLAS Run-1 and 79.8 fb$^{-1}$ Run-2 data 
yield a sensitivity of 5.4\,(5.5) st. dev.~\cite{Aaboud:2636066}.
Stage-1 simplified template cross-sections times $H\rightarrow  bb$ branching were reported~\cite{Aaboud:2666555}.
Dedicated CMS searches in ttH, VBF, ggH, and VH production modes led to
5.6 (5.5) st. dev. and $\mu = 1.04 \pm 0.20$,
where the signal strength $\mu$ is defined as the ratio of the
measured cross-section 
to the SM prediction~\cite{Sirunyan:2636067}.

\vspace*{-1mm}
\subsection{$ttH$}
\vspace*{-1.5mm}
$ttH$ observation in ATLAS has 5.8 (4.9) st. dev.~\cite{Aaboud:2621167} and in
CMS has 5.2 (4.2) st. dev.~\cite{Sirunyan:2636067}.
The full Run-2 139\,fb$^{-1}$ data-set was analysed for $ttH(H\rightarrow \gamma\gamma$)~\cite{ATLAS-CONF-2019-004}.
Separate event selections are applied corresponding to the decay modes of the top quarks (hadronic and leptonic).
Templates from top mass distributions were constructed in $tt\gamma\gamma$, $\gamma\gamma$+jets, and $ttH$ simulations in order to  decompose the continuum background by a template fit to the data.

The CMS observation with 2016 data was 5.2 (4.2) st. dev. (combined $bb$, multilepton, $\gamma\gamma$, $ZZ$ channels)~\cite{Sirunyan:2312113},
including 2017 data, the analysis of  $ttH(H\rightarrow \gamma\gamma)$ resulted in  $\mu=1.7_{-0.5}^{+0.6}$~\cite{CMS-PAS-HIG-18-018}.

In the $ttH$ multilepton channels with $\ell  = e$ or $\mu$, and $\tau$\,(hadronic decay) 
$\mu = 0.96_{-0.31}^{+0.34} (1.00_{-0.27}^{+0.30})$ for 35.9 fb$^{-1}$ (2016 CMS data) was obtained~\cite{CMS-PAS-HIG-18-018}.
For  41.5\,fb$^{-1}$ (2017 ATLAS data)
3.2 (4.0) st. dev.  was achieved~\cite{Aaboud:2299050}.
In an analysis of $ttH$ and $ttW$ production in multilepton final states using 80\,fb$^{-1}$, ATLAS measured the $ttH$ signal 
with 1.8 (3.1) st. dev. above the SM background~\cite{ATLAS-CONF-2019-045}.
 
The  $ttH(H\rightarrow bb$) fully hadronic, single-lepton and double-lepton final states were analysed by CMS, leading to 
3.7 (2.6) st. dev.~\cite{CMS-PAS-HIG-18-030}.

\subsection{$H\rightarrow \mu\mu$}
As the SM $B(H\rightarrow\mu\mu) = 0.022\%$ is very small, currently, only
 limits are set at 95\% CL on $\mu = \sigma(pp\rightarrow H) \times B(H\rightarrow \mu\mu) / \sigma(pp\rightarrow H)_{\rm SM} \times B(H\rightarrow \mu\mu)_{\rm SM}$ 
< 2.1 (2.0) by ATLAS~\cite{ATLAS-CONF-2018-026} and 
< 2.9 (2.2) by CMS~\cite{Sirunyan:2631546}.

\section{Simplified template cross-sections STXS}
STXS was proposed at the Les Houches’15 workshop and by the LHC Higgs boson cross-section working group. The goal was to have a common format for ATLAS, CMS and theory, in particular,
to measure cross-sections per production modes (ggF, VBF, VH, ttH) in different phase space, signal templates $p_{\rm T}(H),~p_{\rm T}(V)$, etc., reducing model dependency and maximizing sensitivity to BSM effects, and
to combine different decay channels in order to increase sensitivity.

An ATLAS combination of the main channels was performed with STXS stage-1~\cite{Aad:2688596}.
In STXS, several channels contribute to different kinematic regions of the same production mode, 
e.g., VH dominated by $H\rightarrow bb$ in high $p_{\rm T}(V)$, while $gg$ and $ZZ^*$ are relevant at low $p_{\rm T}(V)$.
No significant deviation from SM predictions in any kinematic region was observed, and the p-value with respect to the SM hypothesis is 0.80.

A CMS combination of various production modes and decay channels was performed with 35.9\,fb$^{-1}$~\cite{Sirunyan:2640611}.
A first CMS measurement of STXS stage-1 regions in the diphoton channel was performed, covering the gluon fusion (ggH) and vector boson fusion (VBF) production modes.
For STXS stage-1, ggH and VBF bins using $H\rightarrow \gamma\gamma$ based on 77.4 fb$^{-1}$ were analysed~\cite{CMS-PAS-HIG-18-029}. Ten ggH and three VBF parameters were defined, depending on the number of jets and $p_{\rm T}(H)$. 
For each of the thirteen signal parameters, the
measured cross-section was compared with the SM prediction, and good agreement was obtained.

\section{Differential Higgs boson decay cross-sections}

Differential Higgs boson decay cross-sections have been measured for several modes, e.g.,
$H\rightarrow \gamma\gamma$ with ATLAS data~\cite{ATLAS-CONF-2018-028}
and  $H\rightarrow\gamma\gamma, ZZ, 4\ell, bb$ combined modes with CMS data~\cite{Sirunyan:2651932}.

\section{Rare Higgs boson decays}
Further SM Higgs boson decay modes can 
be in reach with growing LHC data sets.
Current ATLAS limits at 95\% CL are:
\begin{itemize}
\item $\mu = \sigma(pp\rightarrow H) \times B(H\rightarrow Z\gamma) / 
\sigma(pp\rightarrow H) \times B(H\rightarrow Z\gamma)_{SM}< 6.6~(4.4)$~\cite{Aaboud:2276364}, and
\item $\mu = \sigma_{\rm HZ} \times B(H\rightarrow cc) / 
\sigma_{\rm HZ} \times B(H\rightarrow cc)_{SM}<110~(150)$~\cite{Aaboud:2304413}.
\end{itemize}

The four-muon final state is experimentally clean containing only very small SM background~\cite{CMS-PAS-HIG-18-025}, however,
the observed (expected) CMS limits are far from the expected SM rates.
The limits at 95\%\,CL are:
\begin{itemize}
\item $B(H\rightarrow J/\psi J/\psi) < 1.8\cdot 10^{-3}$ (obs), $<(1.8^{+0.2}_{-0.1})\cdot 10^{-3}$ (exp), and
\item $B(H\rightarrow \Upsilon\Upsilon)< 1.4\cdot 10^{-3}$ (obs), $<(1.4 \pm 0.1)\cdot 10^{-3}$ (exp).
\end{itemize}

Anomalous couplings were also searched for~\cite{Sirunyan:2652808}.

\section{Higgs boson decays into invisible particles}
A motivation for this search was given by Patt and Wilczek, “Higgs-field portal into hidden sectors”~\cite{Patt:2006fw}.
There are indirect constraints from coupling fits, and direct constraints from searches for Higgs bosons decaying into invisible particles.
Three separate ATLAS searches were performed: $V$(had)$H$(inv), $Z$(lep)$H$(inv) and VBF $H$(inv), with the result 
$B(H\rightarrow {\rm inv}) < 0.26~(0.17)$ at 95\% CL, assuming SM production cross-section~\cite{ATLAS-CONF-2018-054}.

For CMS, the dominant backgrounds are $Z(\nu\nu)$+jets and $W(\ell\nu)$+jets, extrapolated from \mbox{2-lepton} sideband, and from 1-lepton sideband, respectively. The VBF production channel is the most sensitive one.
2016 VBF-only data led to  $B(H\rightarrow {\rm inv}) < 0.33$ (0.25) at 95\% CL.

About 25\% improvement in sensitivity is obtained by adding VH and ggH channels, thus
$B(H\rightarrow {\rm inv}) < 0.26$ (0.20) at 95\% CL (13\,TeV data), and
$B(H\rightarrow {\rm inv}) < 0.19$ (0.15) using 7, 8, 13\,TeV data~\cite{Sirunyan:2638810}.
$ttH$ limits on invisible decays are also set at 95\% CL
$ B(H\rightarrow {\rm inv}) < 0.46\,(0.48)$~\cite{CMS-PAS-HIG-18-008}.

\section{Combination}
The ATLAS results of the measurements from several production and decay modes were combined  taking their uncertainties into account~\cite{Aad:2688596}. 
The result is expressed as a comparison to the SM expectation $\mu=1.11^{+0.09}_{-0.08}$.
Thus, the combined result is compatible with the expectation from the SM.

\section{Higgs boson production modes (ggF, VBF, VH, ttH)}
All major Higgs boson production modes are observed ($>5$ st. dev.) based on 79.8 fb$^{-1}$ ATLAS data, 
assuming SM branching ratios~\cite{Aad:2688596}:
\begin{itemize}
\item ggF, VBF (6.5 st. dev.), 
\item VH (5.3 st. dev.), and 
\item ttH (5.8 st. dev.).
\end{itemize}

There are only small correlations between production modes, and the
results are consistent with the SM expectations.
Similar results are obtained by a CMS data combination~\cite{Sirunyan:2640611}.

Cross-sections are measured for VBF versus ggF.
Individual and combined decay modes are in agreement with the SM expectation~\cite{Aad:2688596}.
A generic parametrization of the measured couplings with respect to the SM was performed for
 $\lambda(tg)$ contributing through ggF loop, as compared to ttH, and
 $\lambda(\gamma Z)$ contributing to the $H\rightarrow \gamma\gamma$ loop, 
 as compared to $H\rightarrow  ZZ$ decays.
The results are also in agreement with SM expectations.

For 137\,fb$^{-1}$ data in the 
$H\rightarrow ZZ \rightarrow 4\ell$ channel ($\ell = e$ or $\mu$),
the cross-section measurement is $\sigma =2.73_{-0.22}^{+0.23}~({\rm stat})~_{-0.19}^{+0.24}~({\rm syst})$\,fb, 
which is consistent with the SM expectation $2.76 \pm 0.14$\,fb~\cite{CMS-PAS-HIG-19-001}. 
Differential cross-sections versus $p_{\rm T}(H)$, $H$ rapidity, and the number of jets were also studied.

\section{Single-top-Higgs production tH, Higgs boson pair-production HH}

The single-top-Higgs production 
$tH(H \rightarrow WW,ZZ,\tau\tau, bb$) was searched for.
A combination with $ttH$ and $H \rightarrow \gamma\gamma$ could give 
sensitivity to the absolute values of the top quark Yukawa coupling and
the Higgs boson coupling to vector bosons $g_{HVV}$, and uniquely, to their relative sign.
The SM-like signal favours a $\kappa_t = 1.0$ over $\kappa_t = -1.0$ by $> 1.5$ st. dev.~\cite{Aad:2677418}.
ATLAS sets a limit on $HH \rightarrow bbbb, bb\tau\tau, bb\gamma\gamma < 6.7~(10.4) \times {\rm SM}$  at 95\% CL.
CMS obtained a limit $HH \rightarrow bbVV, bbbb, bb\tau\tau, bb\gamma\gamma < 22.2~(12.8) \times {\rm SM}$  at 95\% CL~\cite{Sirunyan:2648884}.

\section{Relation of coupling and fermion mass}

\begin{figure}
\includegraphics[width=0.49\linewidth]{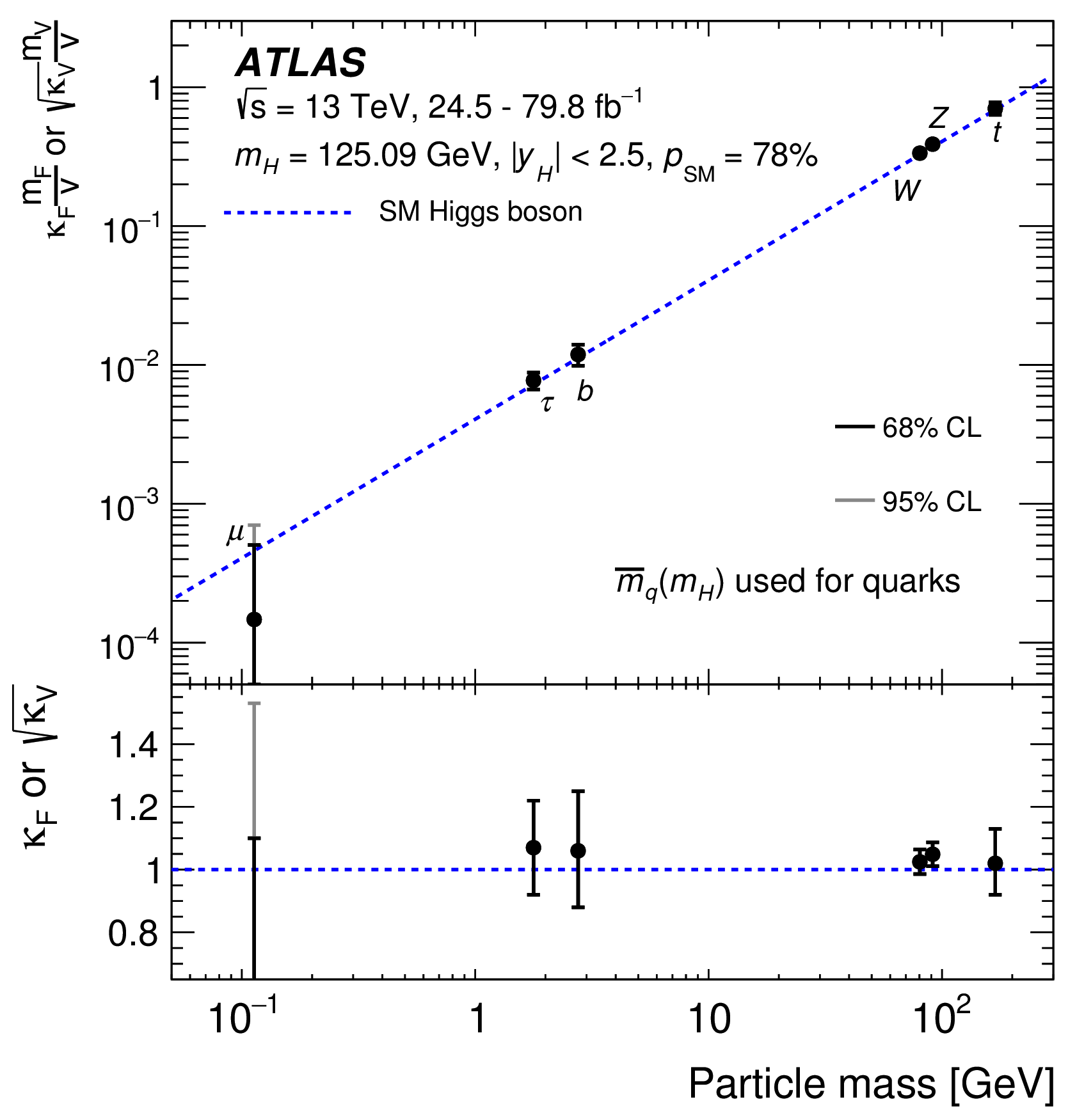}\hfill
\includegraphics[width=0.49\linewidth]{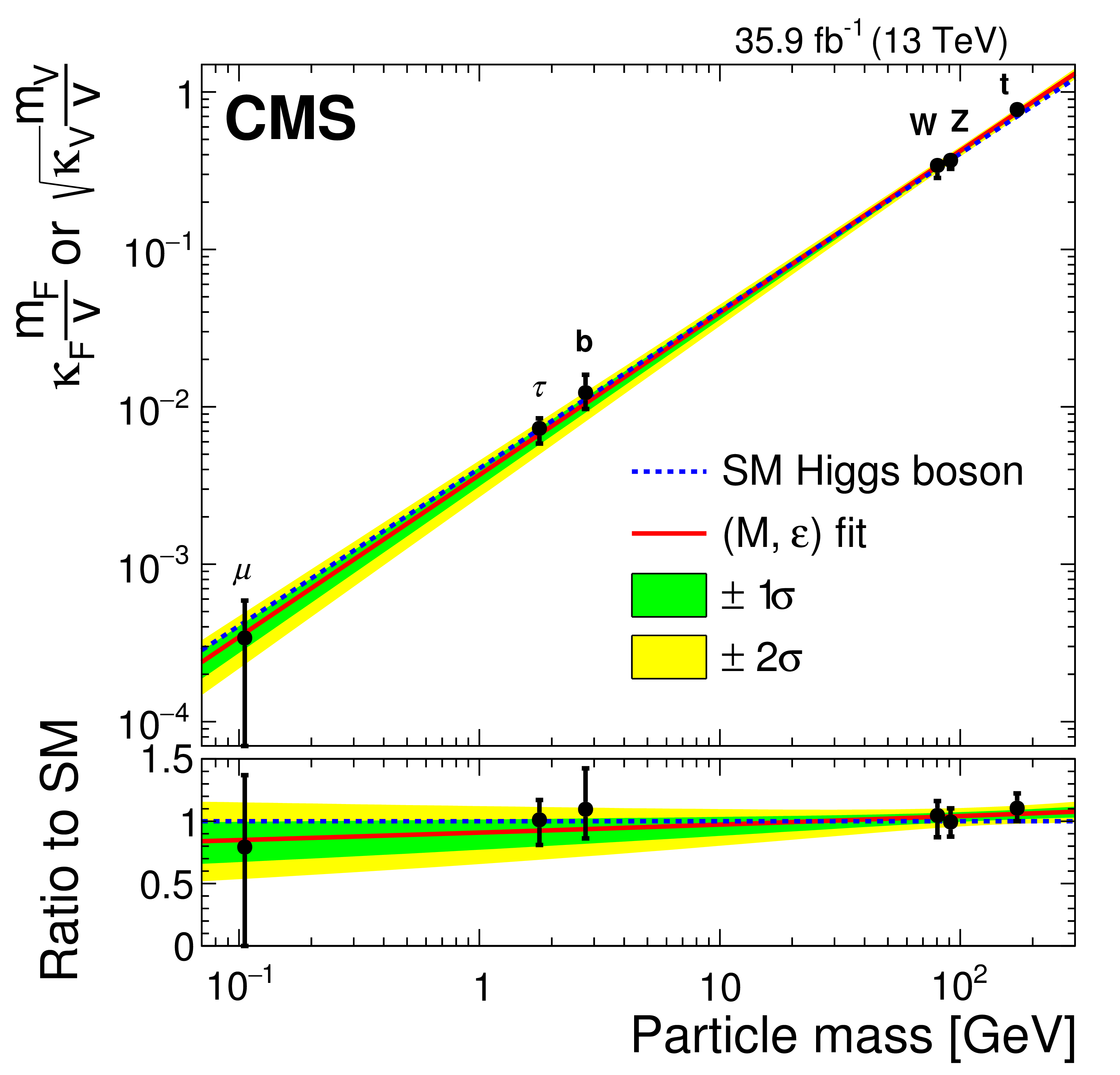}
\caption{
Left (ATLAS result)~\cite{Aad:2688596}:
Reduced coupling-strength modifiers $\kappa_Fm_F/v$ for fermions ($F=t, b, \tau, \mu$) and $\sqrt{\kappa_V}m_V/v$ for weak gauge bosons ($V=W, Z$) as a function of their masses $m_F$ and $m_V$, respectively, and the vacuum expectation value of the Higgs field $v=246$\,GeV. The SM prediction for both cases is also shown (dotted line). The black error bars represent 68\% CL intervals for the measured parameters. For $\kappa_\mu$ the light error bars indicate the 95\% CL interval. The coupling modifiers $\kappa_F$ and $\kappa_V$ are measured assuming no BSM contributions to the Higgs boson decays, and assuming the SM structure of loop processes such as $gg\rightarrow H, H \rightarrow \gamma\gamma$ and $H\rightarrow gg$. The lower panel shows the ratios of the values to their SM predictions. 
Right (CMS result)~\cite{Sirunyan:2640611}:
Result of the phenomenological $(M,\epsilon)$ fit is also overlayed with the resolved $\kappa$-framework model.
\label{fig:coupling}
}
\end{figure}

An interpretation of the results is given in the $\kappa$-framework as a function of the particle mass, assuming only SM contributions to the total width~\cite{Aad:2688596,Sirunyan:2640611}. Figure~\ref{fig:coupling}
shows the linearity as expected from the Higgs boson theory for ATLAS and CMS.
Under the assumption that new physics affects only the Higgs boson self-coupling, the best fit value of
the coupling modifier is $\kappa_\lambda = 4.6^{+3.2}_{-3.8}$, excluding values outside the interval $-2.3 < \kappa_\lambda < 10.3$ at 95\% CL,
while the expected excluded range assuming the SM predictions is 
$-5.1 < \kappa_\lambda < 11.2$~\cite{ATLAS-CONF-2019-049}.

\clearpage
\section{Conclusions and Outlook}
In conclusion, LHC Run-2 operation was very successful.
Observations of the Higgs to tau coupling, 
Higgs to bottom coupling, and
Higgs to top coupling
were reported.
Sensitivity to the  Higgs to muon coupling is approaching.
After establishing these inclusive measurements,
the focus is now on differential measurements and
combinations of all main LHC Higgs boson production modes.
So far, all Higgs boson properties are in agreement with the SM expectations.

The outlook for LHC Run-2 data analysis is
towards a complete analysis of the LHC Run-2 data set (about 140 fb$^{-1}$ per experiment), 
a more detailed understanding of the data to increase the measurement precision, and
a combination of ATLAS and CMS results to increase sensitivities further.
LHC Run-3 is anticipated to add 300 fb$^{-1}$ (during the 2021 to 2023 data-taking), and HL-LHC is approved for 3000 fb$^{-1}$  (expected to start delivering data in 2026) for a new era of measurements with higher precision.
Overall there is a strong and approved LHC programme for new discoveries and further precision measurements.

\section*{Acknowledgments}
I would like the thank the colleagues from the ATLAS and CMS Higgs working groups, and the theorists and phenomenologists present at the conference for the fruitful discussions, as well as the organizers of FFK2019 for their invitation and hospitality. The project is supported by the Ministry of Education, Youth and Sports of the Czech Republic under project number LTT17018.

\bibliographystyle{JHEP}
\bibliography{biblio}

\end{document}